# Entropic signatures of the skyrmion lattice phase in MnSi$_{1-x}$Al$_x$ and Fe$_{1-y}$Co$_y$Si.

C. Dhital[1*] and J. F. DiTusa[2]

[1]Department of Physics, Kennesaw State University, Marietta, GA, 30060, USA

[2]Department of Physics and Astronomy, Louisiana State University, Baton Rouge, LA, 70803, USA and The Purdue School of Science, IUPUI, Indianapolis IN, 46202 USA

**Abstract**

The entropic signatures of magnetic phase transitions in the skyrmion lattice host compounds MnSi$_{0.962}$Al$_{0.038}$ and Fe$_{0.7}$Co$_{0.3}$Si were investigated through low field magnetization and ac susceptibility measurements. These data indicate that the conical to skyrmion transition that occurs with the application of magnetic field in MnSi$_{0.962}$Al$_{0.038}$ is characterized by clear discontinuity in the magnetic entropy as expected for first order topological phase transition. These same magnetoentropic features are negligibly small in isostructural Fe$_{0.7}$Co$_{0.3}$Si due to the level of chemical substitution related disorder and differences in the spin dynamics (range and timescales). Despite the obvious similarities in the magnetic structures of these two compounds, the transitions between these phases is substantially different indicating a surprising non-universality to the magnetic phase transitions in this class of materials.

**Introduction:**

The skyrmion lattice hosting chiral cubic magnets MnSi, FeGe, Cu$_2$SeO$_3$, and Fe$_{1-x}$Co$_x$Si are rich in magnetic anomalies due to the presence of magnetic fluctuations and precursor phenomena associated with the magnetic transitions between topologically trivial (helical, conical and ferromagnetic) and non-trivial (magnetic skyrmion lattice) phases [1-11]. These anomalies often appear as subtle features in the magnetization. However, in real materials these signatures are often smeared out due to the inherent configurational, thermal and chemical disorder. The overall magnetic phase diagram consisting of the helical, conical, skyrmion lattice or A-phase, and the field polarized phase appear very similar for the materials listed above [11]. Nonetheless, there are some discrepancies that have been reported regarding the precursor phase and the nature of the paramagnetic-to-helical phase transition. Multiple studies on MnSi have indicated the presence of a fluctuating precursor phase that causes the helical magnetic transition to be a first order phase transition following the scenario put forward by Brazovskii to describe liquid crystals [12]. With the application of a magnetic field, the transition becomes second order indicating the presence of a tricritical point [12]. However, the existence of this tricritical point is still in question [4]. The universality of this description for all of the magnetic *B20* materials is also a topic of recent debate as it is not clear if the paramagnetic to helimagnetic phase transition in chemically disordered $Fe_{1-y}Co_ySi$ can be described in the same manner with a chiral precursor phase of strongly fluctuating moments [8,10,13]. The signatures of the phase transitions can also be smeared due to substitutional or thermal disorder [8,12,14-16].

One defining feature of the magnetic phase diagram of these compounds is the sequential transition of the helimagnetic, conical, skyrmion lattice and the field polarized phases with the magnetic field [11]. The interplay of a sizeable Dzylloshinskii-Moriya interaction ($D$) (allowed by low symmetry of the crystal structure) and the usual exchange interaction ($J$) cause helimagnetic order in zero field with a wavelength $\lambda \sim J/D$ [1, 11]. The transitions between different magnetic phases in these materials involve both continuous (helical to conical and conical to field polarized) and non-continuous (conical to skyrmion lattice and vice versa) transitions. One method for characterizing these phase transitions is by the calculation of the magnetic entropy change during the transitions. For example, a conventional second order field driven transition to a field polarized state results in negative magnetic entropy change due to a reduction in spin disorder or spin configurational entropy with the application of high magnetic fields [6,17]. With careful magnetization versus temperature measurements, the entropic signatures of magnetic phase transitions can be obtained as evidenced, for example, by previous studies on FeGe [18], $FePd_{1-x}Pt_xMo_3N$ [19] and $GaMo_4Se_8$ [20]. These previous works indicate that the measurement of entropy change through systematic measurements of the magnetic susceptibility can be very powerful tool for identifying and distinguishing between the continuous phase transitions and the transition between topologically trivial and non-trivial phases. Knowledge of the magnetic entropy change can also provide valuable information about the entropy limited topological protection and the decay of magnetic skyrmions [21]. In the current work, we have investigated the magnetoentropic signatures of the phase transitions in two skyrmion hosting systems $MnSi_{0.962}Al_{0.038}$ and $Fe_{0.7}Co_{0.3}Si$. The choice of these two compounds comes from their similar value of transition temperatures combined with their very different levels of chemical substitution. This study allows us to compare the effect of chemical disorder on the magnetic entropy while retaining the same thermal disorder. Our results indicate that the low field entropic signatures are clearly visible in MnSi$_{0.962}$Al$_{0.038}$ whereas these signatures in Fe$_{0.7}$Co$_{0.3}$Si are not distinguishable likely due to the large chemical disorder and the related difference in characteristic correlation lengths [8].

## 2. Experimental Details:

Polycrystalline samples were obtained by arc melting high purity elements in an inert argon atmosphere. These samples were annealed at 1000 $^0$C for approximately 3 days in sealed quartz tubes. Single crystals of MnSi$_{1-x}$Al$_x$ ($x$~0.04) and Fe$_{1-y}$Co$_y$Si ($y$=0.3) were obtained by loading the polycrystalline pellets inside graphite tubes and employing the modified Bridgman method in a RF furnace under a flowing argon environment. The phase purity and crystallinity of our samples were determined by powder and single crystal X-ray diffraction. The concentration of Al in MnSi$_{1-x}$Al$_x$ was determined using Wavelength Dispersive Spectroscopy (WDS) technique in JEOL JSX-8230 Electron Microprobe. For WDS measurement, a small piece of single crystal was taken and finely polished. Then the WDS spectra were taken at multiple spots and the ratio of atomic percent of aluminum to silicon was calculated. The mean value of this ratio was found to be 0.038 with maximum spread of 0.003 from the mean value.

The concentration of cobalt in Fe$_{1-y}$Co$_y$Si was estimated from the comparison of value of magnetic transition temperature, $T_C$, to previously published literature [22-25] which was consistent with the nominal concentration used. Magnetization measurements, both *ac* and *dc*, were carried out in a Quantum Design 7-T MPMS SQUID magnetometer (SCM5) at the National High Magnetic Field Laboratory in Tallahassee, Florida. Unless otherwise stated, the magnetization, *M* vs temperature, *T* data were collected during warming under zero field cooled (ZFC) conditions. The magnetic field was applied along an arbitrary (unknown) direction with respect to the crystallographic axes. For the estimation of entropy and the field variation of ac susceptibility, the data were corrected for demagnetizing field and internal fields are reported. For the demagnetization correction of magnetic field, the relation [26, 27] $H=H_a-NM$, where *H* is the internal field, $H_a$ is the applied field, *N* is the demagnetization correction and *M* is the magnetization at that field, was used. The MnSi$_{0.962}$Al$_{0.038}$ sample employed for these measurements was nearly rectangular in shape with dimensions of 3 x 2 x 0.9 mm. The magnetic field was applied parallel to the longest side of the sample for all measurements. A demagnetizing factor $N=$ 0.17 was estimated for this sample and the internal field *H* was calculated using the relation $H=H_a-NM$ [26]. For $Fe_{0.7}Co_{0.3}Si$, the sample dimensions were 1.95 x 1.95 x 1.93 mm. The internal field was obtained using the estimated demagnetizing factor of $N=0.33$. A complete description of demagnetization correction procedure for ac susceptibility and dc fields is presented in the supplementary materials [28].

In the following sections, we discuss systematic measurements of the low field magnetization to map the magnetoentropic signatures of the phase transitions in $MnSi_{0.962}Al_{0.038}$ and $Fe_{0.7}Co_{0.3}Si$.

## 3. Results and Discussion:

### a. $MnSi_{1-x}Al_x$.

Nominally pure MnSi is an itinerant magnet with magnetic transition temperature $T_C=29$ K and an ordered moment of ~0.4 $\mu_B$ per formula unit. The zero-field ordered magnetic state is helimagnetic with a pitch length ($\lambda$) of ~180 Å which propagates along the [111] direction of its cubic crystal structure. Just below $T_C$, the magnetic state evolves with magnetic field through helical-conical-skyrmion (or A-phase)-conical and field polarized phases [11, 15, 29-32]. Partial Al substitution for Si creates a negative chemical pressure on the lattice increasing ordered magnetic moment to ~0.5 $\mu_B$ and $T_C$ to ~39 K for $x=0.038$ with all qualitative features of the phase diagram remaining same as that of nominally pure MnSi. A complete magnetic phase diagram of $MnSi_{0.962}Al_{0.038}$ is published in our previous work [30]. Figure 1 shows the *dc* magnetization as a function of temperature along with the *ac* magnetic susceptibility as a function of internal magnetic field for $MnSi_{0.962}Al_{0.038}$. Figure 1(a) presents the magnetization, $M$, and its temperature derivative, $dM/dT$ as function of $T$ measured at applied field of 0.01 T. A close examination of $dM/dT$ indicates two inflection points (kinks) representing the transition to the helical phase at $T_C$ ~39 K and a transition from the paramagnetic to the fluctuating disordered precursor phase $T_F$ ~ 43 K. The shaded region indicates the precursor phase with fluctuating moments, similar to what is observed in nominally pure MnSi [6,7,12]. Figures 1b and 1c present the real ($\chi'$) and imaginary ($\chi''$) parts of the *ac* susceptibility as a function of field measured at 20 Hz, 100 Hz and 500 Hz at $T=37.5$ K. The susceptibility resembles that of a typical skyrmion hosting cubic magnet [11, 29-32]. There are four critical magnetic fields identified as $H_{C1}$, $H_{A1}$, $H_{A2}$, and $H_{C2}$. These fields correspond to continuous transition from a multidomain helical phase to a single domain conical phase at $H_{C1}$, a conical to skyrmion lattice or *A*-phase at $H_{A1}$, the *A*-phase to conical phase transition at $H_{A2}$, and the conical to field polarized phase at $H_{C2}$. In the region between $H_{A1}$ and $H_{A2}$ there is reduction in real part of *ac* susceptibility indicating the slow movement of massive magnetic objects. Furthermore, the entry and exit of the *A*-phase are characterized by a peak in $\chi''$ indicating a finite dissipation process during the phase transition between the topologically trivial conical phase and the topologically non-trivial skyrmion phase and vice versa [11].

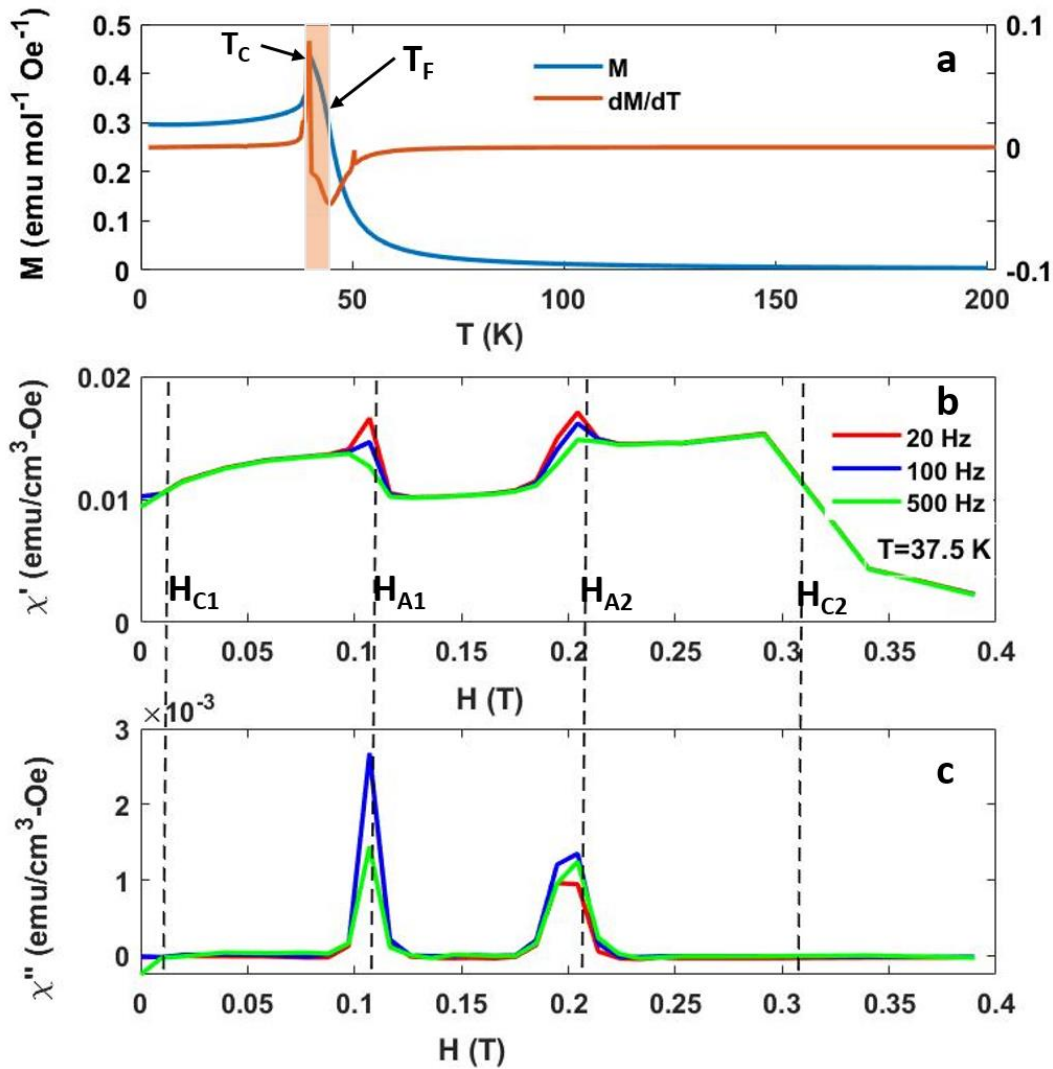

**Fig.1** *dc* and *ac* magnetic properties of MnSi$_{0.962}$Al$_{0.038}$. (a) *dc* magnetization, *M* and its derivative, *dM/dT* as function of temperature, *T* measured at an applied magnetic field of 0.01 T. (b) Real part of the *ac* susceptibility, $\chi'$ as function of internal field, *H*, at several frequencies at *T*=37.5 K. (c) Imaginary part of the *ac* susceptibility $\chi''$ as function of *H* at the same frequencies and temperature.

The next important feature in the *ac* susceptibility is the frequency dependence of the peak height (maxima) at the entry and exit point of the *A* phase. The amplitude of peak is symmetric for $\chi'$ whereas the amplitude of peak asymmetric for $\chi''$. The decrease in the peak amplitude in $\chi'$ with an increase in frequency is consistent with nominally pure MnSi indicating a slow relaxation process involving the skyrmion lattice [11, 29]. However, the frequency dependence of $\chi''$ is non monotonic. The peak amplitude of $\chi''$ becomes maximum at 100 Hz indicating the characteristic frequency of 100 Hz in this system. This characteristic frequency is less than that compared to nominally pure MnSi (~1 kHz) [29] but is consistent with decrease in characteristic frequency in chemically substituted disordered systems $Mn_{1-x}Fe_xSi$ [29] and $Fe_{1-y}Co_ySi$.

After characterizing the magnetic behavior near the transition to the *A*-phase, we turn to identifying the entropic signatures of the magnetic phase transitions. For this analysis, a large number of constant field *M* vs. *T* measurements were taken between 0.005 T and 0.4 T over a temperature range of 30 K to 52 K. The sample was cooled in zero field from 55 K to 30 K followed by the application of the magnetic field at 30 K. *M* vs *T* data were collected during warming. The results of this procedure are presented in Fig. 2. Fig. 2a presents the *M(T)* for each field where data were taken. The magnetic transition from the helimagnetic state to the helical fluctuating state is indicated by kinks in Fig. 2a. Fig. 2b presents the temperature derivative, *dM/dT* where this transition is more distinct. The peaks in *dM/dT* represent transitions between the different magnetic states. The change in magnetic entropy is calculated using the standard equation derived from a Maxwell relation: $\Delta S_M = \int_0^H \left(\frac{\partial M}{\partial T}\right)_H dH$ and is presented in Fig. 2c. For clarity the curves in Fig.2b and Fig. 2c are offset as indicated in the figure caption. Note here *H* refers to the internal field after demagnetization correction using relation: $H=H_a-NM$.

According to Maxwell's thermodynamic identity: $\left(\frac{\partial S}{\partial H}\right)_T = \left(\frac{\partial M}{\partial T}\right)_H$, the temperature derivative *dM/dT* and hence *dS/dH* can be viewed as thermodynamic variables that give information complementary to the more traditional measurement of the heat capacity *C = T(dS/dT)*. Peaks and valleys in *dS/dH* indicate field-driven phase transitions and ultimately can indicate the entropy changes associated with the transitions. To better represent the different magnetic transitions the values of *dS/dH* and the entropy change *ΔS$_M$* , determined by integrating the data numerically, are presented as color map in Fig. 3 a and b respectively. From Fig. 3a, it is clear that in the high field region *dS/dH* is negative (black) indicating a reduced magnetic entropy due to the reduction in spin orientational disorder as the system reaches the field polarized state. At lower field, however, there are ridges (positive entropy changes) and valleys (negative entropy changes) indicative of the different magnetic phases. The entropy change, $\Delta S_M$, determined by integrating $dM/dT$ ($=dS/dH$) over the field $H$ is plotted in Fig. 3b. The entropy appears to increase in the paramagnetic region at high fields, this is primarily due to increase in $dM/dT$ with $T$ caused by paramagnetic spin fluctuations.

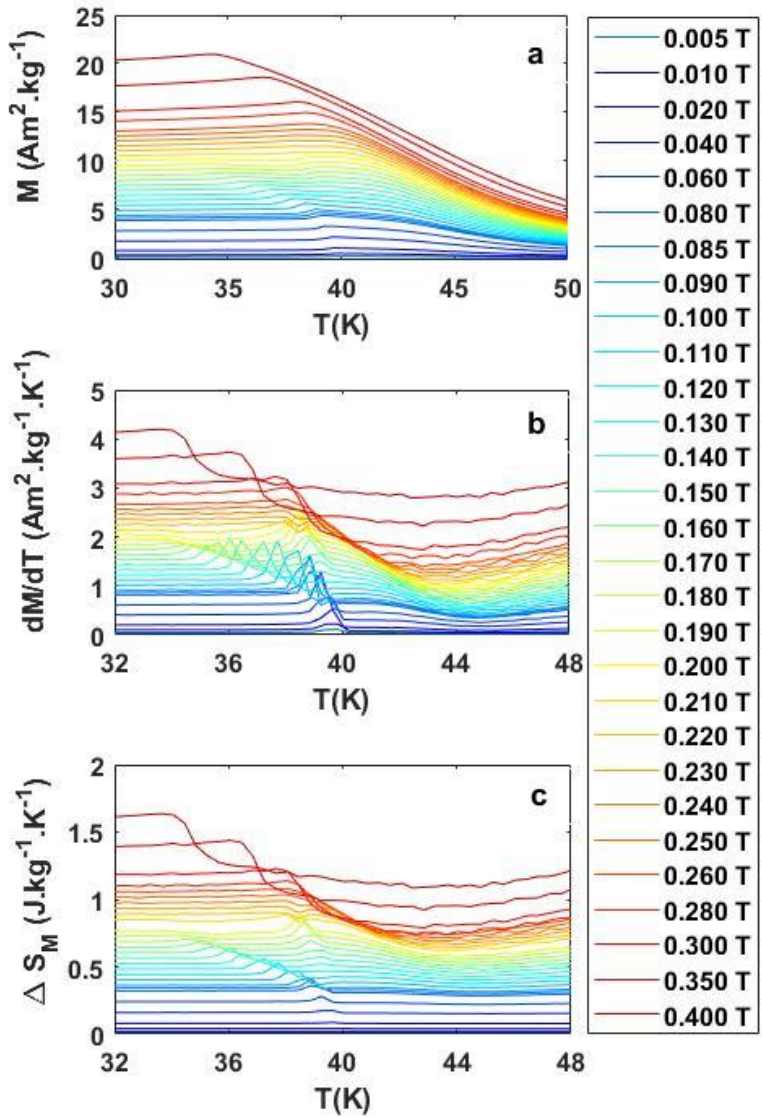

Fig. 2 Magnetic entropy changes in MnSi$_{0.962}$Al$_{0.038}$. (a) Magnetization, $M$, as function of temperature, $T$ for a range of applied magnetic field between $H_a= 0.005$ and $H_a = 0.4$ T. (b) The temperature derivative, $dM/dT$, of the data shown in (a). For clarity these curves are offset by 0.1 A m$^2$ kg$^{-1}$ K$^{-1}$ for every 0.01 T field increment. (c) The change in magnetic entropy $\Delta S_M$. For clarity the curves are offset by 0.04 J kg$^{-1}$ K$^{-1}$ for every 0.01 T field increment. The entropy data are obtained using internal field, $H$, after demagnetization correction.

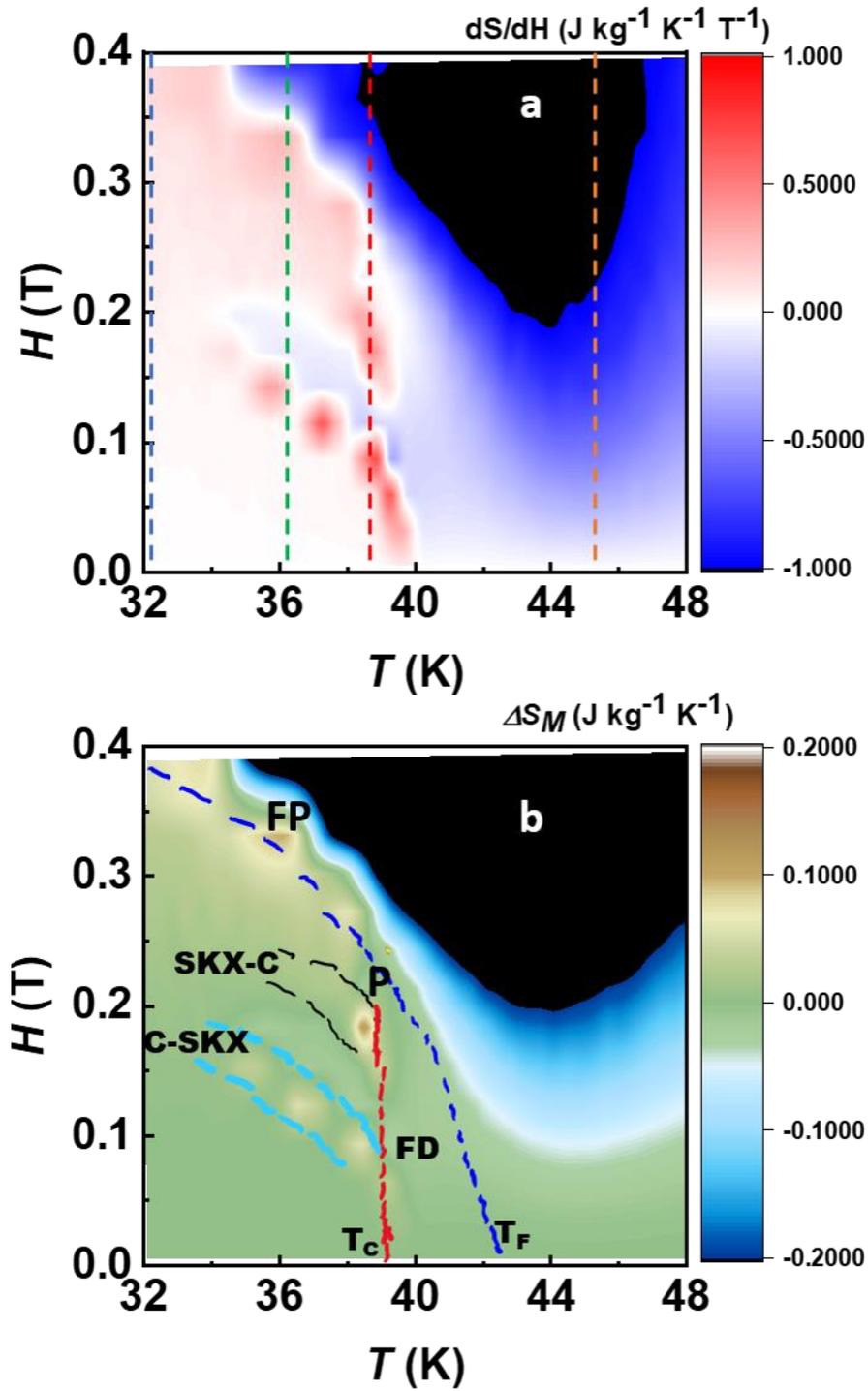

Fig.3 Contour maps of the magnetic entropy for MnSi$_{0.962}$Al$_{0.038}$. (a) Color map of the temperature, $T$, derivative of the magnetization, $M$, $(dM/dT)_H = (dS/dH)_T$ where $S$ is the magnetic entropy and $H$ is the internal magnetic field. The vertical dashed lines represent the cuts along $H$ that are plotted in Fig. 4. (b) Entropy change $\Delta S_M$ as function of $T$ and $H$ determined by numerically integrating the data in (a). The regions of different magnetic structures are distinguished. *FP*: Field polarized, *C*: Conical magnetic state, *FD*: Fluctuating disordered state, *A*: Skyrmion lattice or *A*-phase, $T_C$: magnetic transition temperature from the fluctuating disordered state to ordered state at $H=0$. $T_F$: Transition from the paramagnetic to fluctuating disordered state, *P*: tricritical point separating the first order transition at low field to a continuous change at high field.

The nearly sharp vertical feature around 39 K in Fig. 3a and Fig. 3b delimits the first order phase transition between the fluctuating disordered (*FD*) and the magnetically ordered state similar to that observed in nominally pure MnSi [12]. At temperatures below the first order line, two regions of positive entropy are observed at boundaries of the *A*-phase as indicated by the coincidence of peaks in χ' and $dS/dH$ plotted in Fig. S1 of supplementary materials [30]. The entropy has a discontinuous jump at those boundaries indicating the first order transition from the conical to the SKX phase and vice versa.

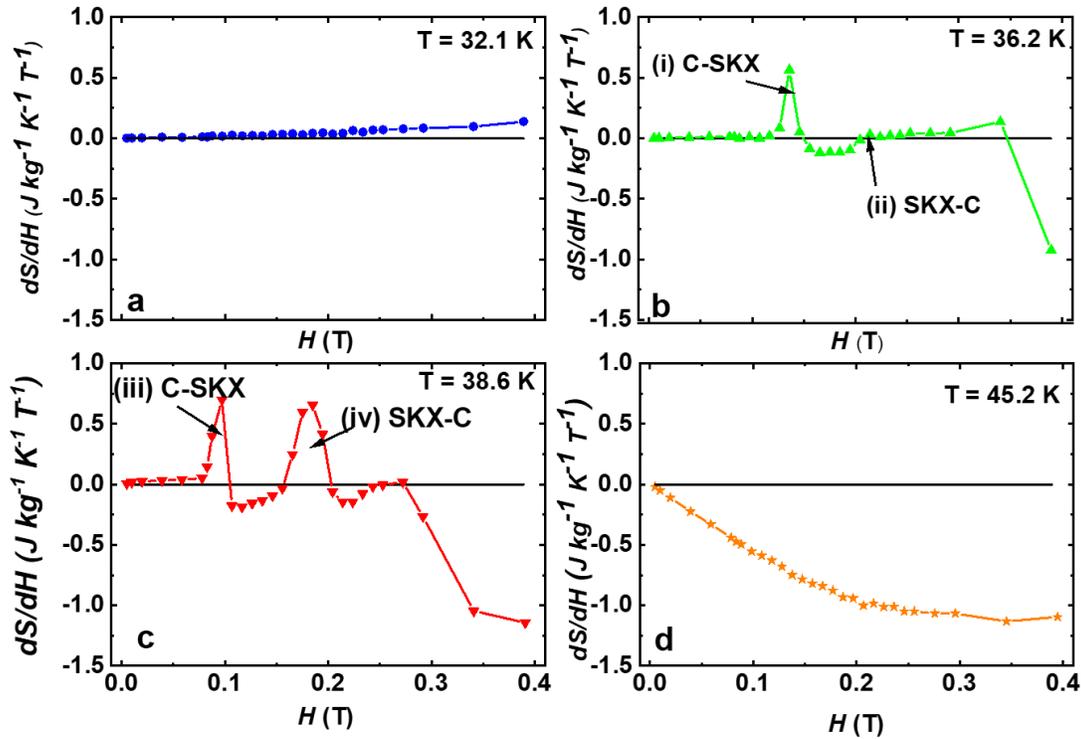

Fig. 4: Rate of change of magnetic entropy in MnSi$_{0.962}$Al$_{0.038}$ obtained from the constant temperature, $T$, cuts (i.e. along the internal magnetic field, $H$) in Fig. 3a at (a) 32.1 K, (b) 36.2 K, (c) 38.6 K and (d) 45.2 K. The peaks and valleys indicate magnetic phase transitions: (i) Conical to skyrmion lattice (C-SKX) (ii) Skyrmion lattice to conical (SKX-C) at 38.6 K, (iii) conical to skyrmion (C-SKX) at 38.6 K (iv) Skyrmion to Conical (SKX-C) at 38.6 K. The black horizontal line represents the zero-reference line.

We have also quantified the entropy associated with the magnetic phase transitions. Fig. 4 presents cuts of the data along $H$, for 4 different temperatures as shown in Fig. 3a. At 32.1 K there is only a slight increase in $dS/dH$ indicating no distinct entropy associated with the helical to conical transition. $dS/dH$ is also slightly positive in conical region just below conical- field polarized transition line $H_{C2}$. Although there is no distinct peak, this could be due to the presence of a fluctuating disordered phase near conical-field polarized phase boundary as observed in FeGe [18]. $dS/dH$ ultimately becomes negative (not shown) at higher fields as the system enters the field polarized state. The shape of curve at 36.2 K (green) is more interesting. As the field is increased, the entropy increases sharply around $H_{A1}$~0.11 T indicating sudden absorption of heat during conical to SKX phase transition. The heat is immediately released upon increasing the field further as indicated by the sudden drop to a negative value of the entropy change. Upon further increasing the field, there is another small increase in entropy at $H_{A2}$ ~0.21 T. This corresponds to SKX to conical phase transition. At 38.2 K, just below $T_C$, (red cure in Fig. 4(c)) $dS/dH$ displays even more interesting behavior and reveals the true nature of first order phase transition from SKX to conical and vice versa. Upon increasing the field, the entropy change discontinuously acquires large positive value at $H_{A1}$~0.09 T and immediately drops to negative value. Upon further increasing the field, there is another similar discontinuity that begins around $H_{A2}$~0.18 T (See comparison of $dS/dH$ and $\chi'$ in supplementary materials). Finally, the red curve becomes continuously negative in the field polarized state. The shape of the curves at 36.2 and 38.6 K indicate that the transition from the topologically non-trivial (skyrmion lattice) and the topologically trivial (conical) is a first order transition with a discontinuous change in magnetic entropy. The difference in the discontinuity at $H_{A2}$ at 36.2 K and 38.6 K comes from the fact that the green curve (36.2 K) lies in proximity to the lower temperature boundary of skyrmion pocket [30] while the red curve taken at 38.6 K lies just below $T_C$ showing stronger effect. Fig. 4d displays the entropy change in the paramagnetic regime at $T$=45.1 K (orange curve) where only a negative $dS/dH$ is observed with field consistent with the idea that the field tends to polarize the $H$=0 paramagnetic state. We have also quantified the entropy change and the associated heat exchange during the transition as determined from the data in Fig. 4. The results are presented in Table I.

**TABLE I**. Latent entropies and heats of transitions as determined by integrating the $dS/dH$ curves shown in Fig. 4. (Only peaks that are distinguishable above the base line are included)

| Transition | $\Delta S$ (mJ kg$^{-1}$ K$^{-1}$) | $T$ (K) | $Q$ (mJ kg$^{-1}$) |
|---|---|---|---|
| (i) C-SKX | 6.8 (4) | 36.2 K | 246 (12) |
| (iii) C-SKX | 10.8 (4) | 38.6 K | 417 (10) |
| (iv) SKX-C | 18.1 (2) | 38.6 K | 699 (13) |

**b. Fe$_{1-y}$Co$_y$Si**

In the Fe$_{1-y}$Co$_y$Si substitution series, both end members FeSi and CoSi are non-magnetic. FeSi is a small band gap insulator with a very small thermally activated magnetic susceptibility whereas CoSi is an interesting semimetal thought to host complex multifold Fermions [24-25]. However, the solid solution Fe$_{1-y}$Co$_y$Si shows interesting helimagnetic and magnetic skyrmion lattice behavior over a wide range of $y$ (0.05 < $y$ <0.8) along with insulator to metal transition at low Co concentration, $y$~0.01 [24-25]. The magnetic transition temperature varies from few K up to 60 K depending upon $y$, creating a dome shaped $T_C$ vs $y$ curve. The helix pitch length and propagation direction are also function of $y$ [8, 10, 11, 13, 22-25, 34]. The helix pitch length initially decreases reaching a minimum around $y$~0.3 and then increases up to $y$~0.8. For this investigation, we have chosen Fe$_{0.7}$Co$_{0.3}$Si with $T_C$ = 43 K, close to that of MnSi$_{0.962}$Al$_{0.038}$. The magnetization and *ac* magnetic susceptibility of Fe$_{0.7}$Co$_{0.3}$Si are presented in Fig. 5.

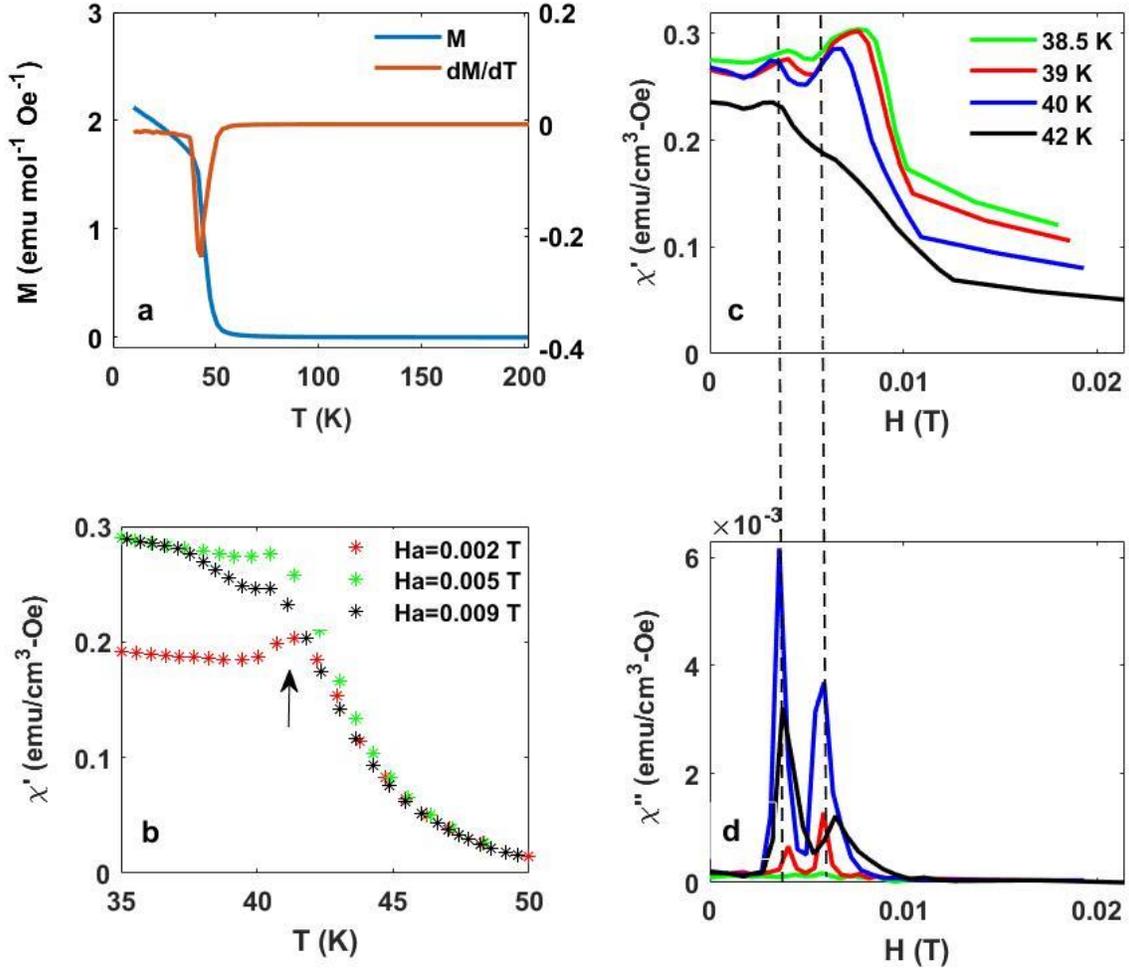

Fig. 5 Magnetic properties of $Fe_{0.7}Co_{0.3}Si$. (a) Magnetization, $M$, vs. temperature, $T$, and $dM/dT$ vs. $T$ at an applied magnetic field of 50 Oe (applied) for a single crystal specimen. (b) Real part of the ac susceptibility, $\chi'$, as function of $T$ for applied fields identified in the figure measured at $f$=20 Hz. (c) Variation of $\chi'$ with $H$ (internal field). (d) Variation of the imaginary part of the ac susceptibility, $\chi''$ with $H$. The vertical lines represent the boundaries of skyrmion lattice phase. The values of $\chi'$ and $\chi''$ are corrected for demagnetizing field as described in supplemental materials [28].

Fig. 5a displays the variation of $M$, and $dM/dT$ as function of $T$ for $Fe_{0.7}Co_{0.3}Si$. Magnetic ordering takes place at $T_C \sim 43$ K which is also reflected in $\chi'$(Fig. 5b). The cusp like feature in the susceptibility is an indication of a typical paramagnetic-to-helimagnetic transition. The shape of the derivative ($dM/dT$) curve is distinct from that of MnSi and $MnSi_{0.962}Al_{0.038}$ indicating that the fluctuation disordered state may be absent or smeared out. Figs. 5c and 5d show how $\chi'$ and $\chi''$ evolve with $H$ indicating the typical signature of the helical, conical, skyrmion lattice and the field polarized phases. The features (peaks in χ') are not as obvious as in MnSi$_{0.962}$Al$_{0.038}$, however χ" is sharply peaked about the entry and exit fields of the *A*-phase. It is not clear if the variations in χ' are smaller in Fe$_{0.7}$Co$_{0.3}$Si because of the presence of chemically induced disorder associated with the doping or if the difference stems from the very large length scale of the skyrmion lattice and the magnetic domains, which are orders of magnitude larger than the length scale expected to be associated with variations in Co density.

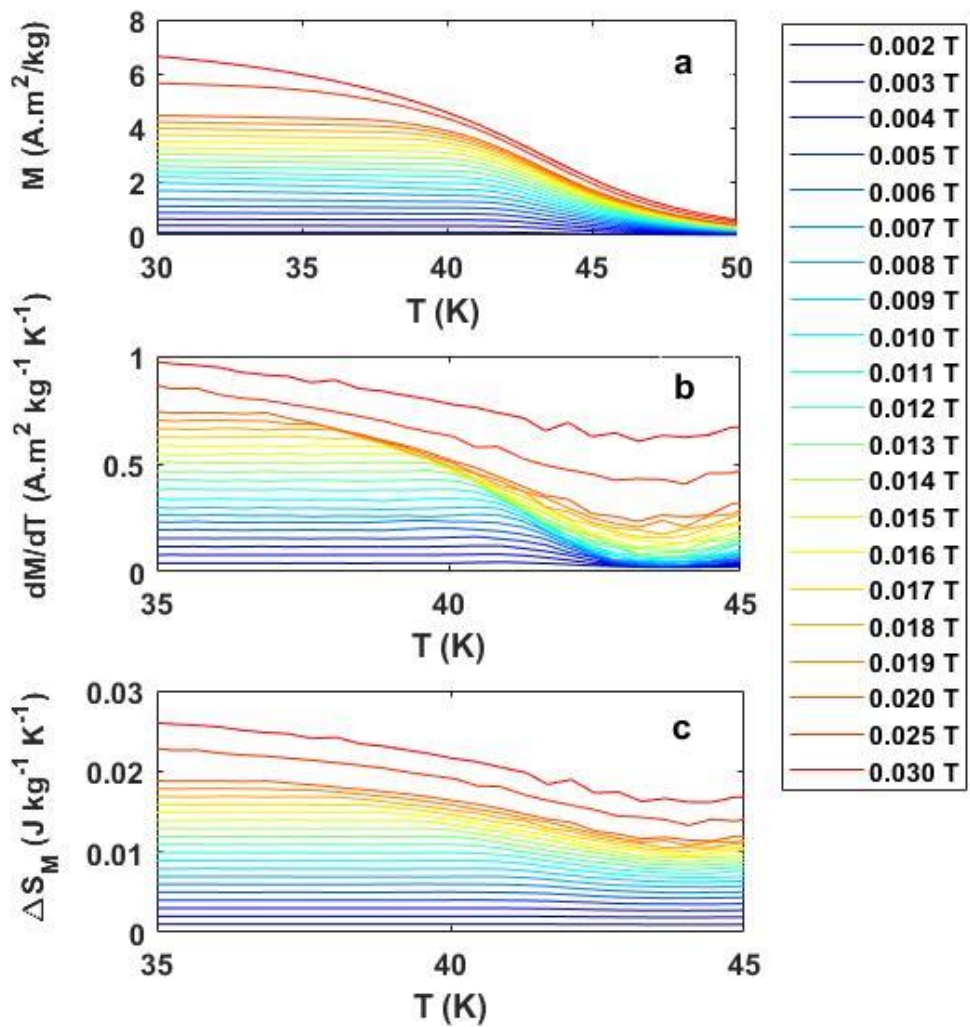

Fig. 6 Magnetocaloric properties of Fe$_{0.7}$Co$_{0.3}$Si. (a) Magnetization, *M* vs temperature, *T*. (b) *dM/dT* vs *T*. For clarity the curves are offset by 0.04 Am$^2$ kg$^{-1}$ K$^{-1}$ for every 0.001 T field increase (c) Change in entropy *ΔS$_M$* vs *T* at various applied fields. For clarity the curves are offset by 0.001 J kg$^{-1}$ K$^{-1}$ for every 0.001 T applied field increase. For the entropy calculation the field values are corrected for demagnetization fields as described in supplemental materials [28]. The applied field is indicated in the key.

Next, we turn to entropic signature of magnetic phase transitions in $Fe_{0.7}Co_{0.3}Si$. For this analysis, *M* vs *T* data were taken in the temperature range of 30 to 52 K during warming under zero field cooled conditions. The temperature derivative and the change in entropy are determined using the same procedure discussed in the previous section. The results are plotted in Fig. 6a, 6b and 6c. The range of *dS/dH* is a factor of ten small than that seen in $MnSi_{0.962}Al_{0.038}$. In addition, the low field features seen in $MnSi_{0.962}Al_{0.038}$ associated with the positive entropy at the boundaries of the skyrmion lattice phase (See supplementary Fig. S2 for comparison of χ' and *dS/dH* for $Fe_{0.7}Co_{0.3}Si$) are not distinguishable. The color map of *dS/dH* and the $\Delta S_M$ are presented in Figs. 7a and Fig 7b respectively. Except for the black regions corresponding to the field polarized region, the map is dominated by small values of *dS/dH* and Δ*S*. In fact in the helical/conical region the entropy seems to be around zero (cyan).This indicates that any subtle magnetoentropic features at low field region are much smaller than in $MnSi_{0.962}Al_{0.038}$. All of the low field entropic signatures are significantly weaker when compared to that of $MnSi_{0.962}Al_{0.038}$ ($M_s$~0.42 $\mu_B$/FU), even when taking into account the slightly smaller size of the magnetic moment in $Fe_{1-y}Co_ySi$ ($M_s$ ~0.3 $\mu_B$/FU).

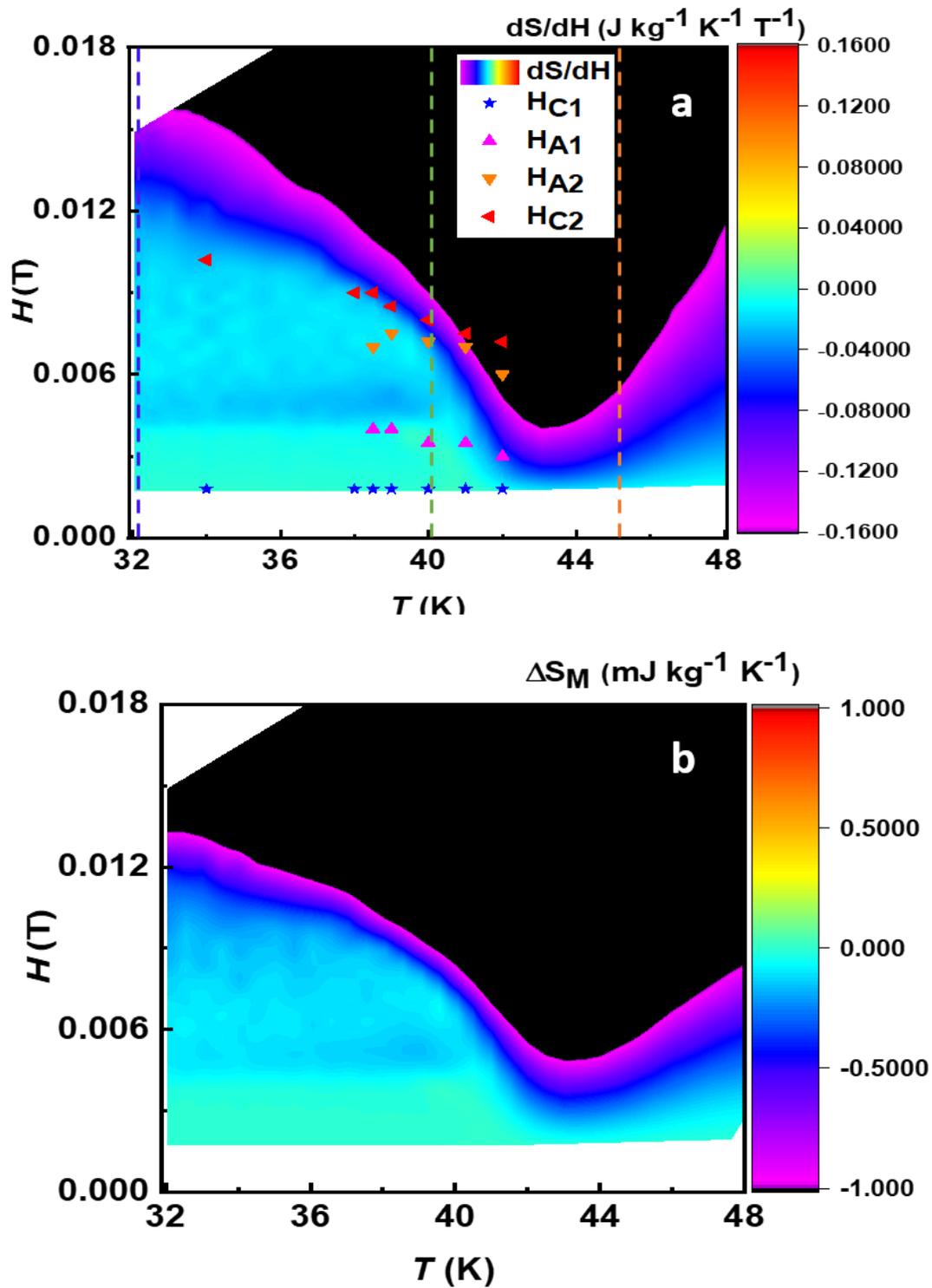

**Fig. 7** Color maps of Fe$_{0.7}$Co$_{0.3}$Si (a) *dS/dH* as function of *T* and *H* (b) $\Delta S_M$ as function of *T* and *H*. Here *H* refers to the internal field after correcting for demagnetization fields as described in supplemental section [28].

The entropy change *dS/dH* at three different temperatures representing the three different regions of the phase diagram are plotted as function of *H* and presented in Fig 8. Unlike the MnSi based system, clear signatures of phase transitions are not visible in the color map and the cuts. The almost horizontal blue curve at 32 K indicates there is no discernable change in magnetic entropy as the system undergoes the helical to conical transition, just as was concluded for MnSi$_{0.962}$Al$_{0.038}$. At 32 K, *dS/dH* ultimately drops to a negative value (not shown) in the field polarized phase at higher values of field. The nature of *dS/dH* at 40 K (green curve) shows some change in the slope at the boundaries of *A*-phase. However, the response is smaller than seen in MnSi$_{0.962}$Al$_{0.038}$ at 38.6 K by nearly a factor of 100. From Fig 8b, the entropy change and the associated latent heat are calculated. The value of entropy change Δ*S* during the conical-to-skyrmion lattice phase change (C-SKX) is estimated to be 0.006 (3) J kg$^{-1}$ K$^{-1}$. Similarly, the heat of transition for (C-SKX) transition is estimated to be 0.24 (5) J kg$^{-1}$. These values are significantly smaller than the values obtained in MnSi$_{0.962}$Al$_{0.038}$ near *T$_C$*. We have compared our estimates of the change in entropy with that of nominally pure MnSi, and FeGe in Table II.

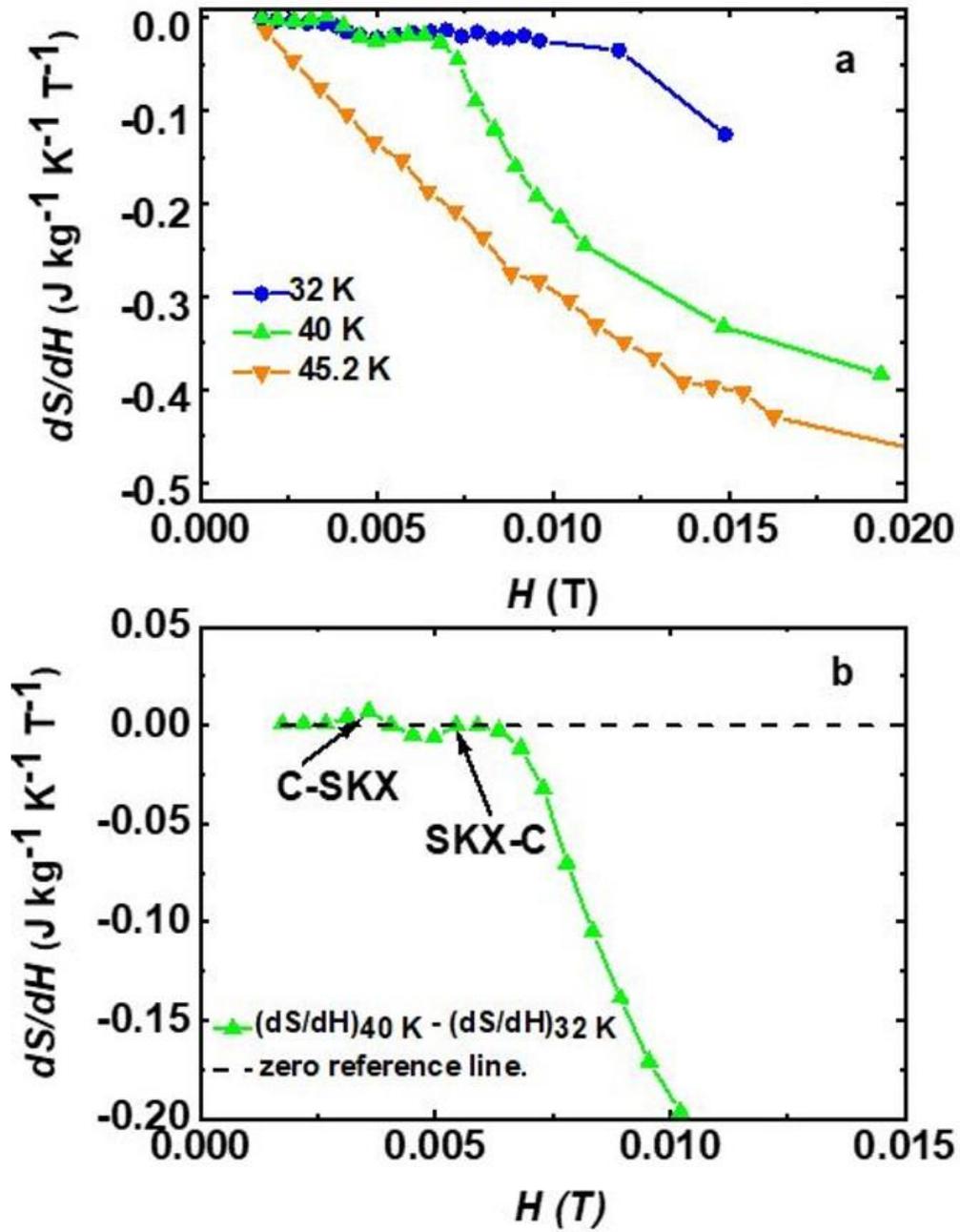

Fig 8. Rate of change of magnetic entropy in $Fe_{0.7}Co_{0.3}Si$ obtained from the cuts along $H$ in Fig. 7a. (a) $dS/dH$ as function of internal field $H$. (b) $dS/dH$ at 40 K.

**Table II**: Comparison of total change in magnetic entropies associated with the transition to the skyrmion lattice phase in MnSi, MnSi$_{0.962}$Al$_{0.038}$, FeGe and Fe$_{0.7}$Co$_{0.3}$Si. (The comparison includes C-SKX phase transition)

| System | $\Delta S_M$ | $\lambda$ | Reference | $\lambda^3 \Delta S_M$ |
|---|---|---|---|---|
| MnSi | 20 mJ kg$^{-1}$K$^{-1}$ | 180 Å | Baur et. al; Phys. Rev. Lett. **110**, 177207 [12] | 6.83x10$^{-22}$ J/K |
| MnSi$_{0.962}$Al$_{0.038}$ | 10.8 mJ kg$^{-1}$K$^{-1}$ | 180 Å | This work | 3.62x10$^{-22}$ J/K |
| FeGe | 0.9 mJ kg$^{-1}$K$^{-1}$ | 690 Å | J.D. Bocarsly et. al; Phys. Rev. B **97**, 100404(R) [15], B. Lebech et al; J. Phys.: Cond. Mat. 1, 6105 [37] | 2.48x10$^{-21}$ J/K |
| Fe$_{0.7}$Co$_{0.3}$Si | 0.006 mJ kg$^{-1}$K$^{-1}$ | 390 Å | This work, S.V. Grigoriev et al; Phys. Rev. Lett. 102, 037204[38] | 2.250x10$^{-24}$ J/K |

**Conclusions**: We have presented magnetic and magnetoentropic signatures of a series of magnetic phases and phase transitions in two cubic magnets MnSi$_{0.962}$Al$_{0.038}$ and Fe$_{0.7}$Co$_{0.3}$Si. As usual, in both systems we find a decrease in magnetic entropy at high fields associated with the field polarized region. The temperature and field range investigated in this work is not large enough to recover the complete magnetic entropy change as the system fully field polarizes. However, the values of the entropy change found in the field polarized region are consistent with previous work when the values of field used are taken into account [34, 35]. Instead, the focus of this work is on the low field region where features associated with the topological transitions are found. Both compounds are itinerant magnets with significant paramagnetic spin fluctuations well above $T_C$ as proposed by Moriya's spin fluctuation theory [36]. In addition, we find no significant change in entropy associated with the helical to conical transition (at $H_{C1}$) or to the conical to field polarized phase (at $H_{C2}$) in both compounds. This indicates continuous transitions at these fields in agreement with previous measurements reported for FeGe [18]. However, at the boundaries of the skyrmion lattice phase we find a clear positive entropy change in MnSi$_{0.962}$Al$_{0.038}$ and very small, but non-zero, changes in Fe$_{0.7}$Co$_{0.3}$Si. In our investigation we have applied a magnetic field in an arbitrary (unknown) direction with respect to the orientation of the single crystals. It has been observed in previous studies [14, 29, 39, 40] that the order of magnetic phase transitions at $H_{C1}$ and $H_{C2}$, the values of critical fields ($H_{C1}$, $H_{C2}$, $H_{A1}$, $H_{A2}$), and the temperature interval for skyrmion phase all depend up on the orientation of the crystal with respect to the field as well as the detailed sequence of changes to the temperature and field and the crystalline disorder. The primary focus of the present work is to investigate the entropic signatures associated to the skyrmion phase and the role of disorder. In our investigation, we observed a discontinuity in entropy only at $H_{A1}$ and $H_{A2}$ indicating the first order nature of the phase transitions. Given the value of $H_{C1}$ that we observe and the absence of an entropy anomaly at either $H_{C1}$ or $H_{C2}$, we point that the magnetic field was oriented closer to the easy axis ([111]) than the hard axis ([001]) of our MnSi$_{0.962}$Al$_{0.038}$ crystal. A full investigation of the entropy changes with respect to field orientation for MnSi would be of interest to elucidate orientation dependence of the magnetic phase transitions.

A comparison of the entropy changes measured for the C-SKX transition in MnSi, FeGe, MnSi$_{0.962}$Al$_{0.038}$, and Fe$_{0.7}$Co$_{0.3}$Si are presented in Table II where variations of more than three orders of magnitude are reported. It is important to recognize that there are large discrepancies in the size of the helical pitch, $\lambda$, and the size of the ordered magnetic moment that may be responsible for some of the variations observed. For example, the helical pitch length ion MnSi is ~18 nm with an ordered magnetic moment of 0.4$\mu_B$, while in FeGe it is 69 nm with an ordered moment of 0.9 $\mu_B$. We have accounted for the size discrepancy by multiplying the measured $\Delta S$ by $\lambda^3$ to represent the entropy change per skyrmion lattice site in Table II. A comparison of $\lambda^3 \Delta S$ for FeGe and MnSi shows similar values with at least some of this difference likely due to the larger ordered moment in FeGe. The modest disorder induced by substituting Al for 3.8% of the Si in MnSi yields a moderate decrease in the size of the entropy change associated with entering the *A*-phase. However, the larger disorder associated with the chemical substitution of Co for Fe in Fe$_{0.7}$Co$_{0.3}$Si almost completely washes out the phase transition with a reduction in the entropic signal of the transition by more than 2 orders of magnitude even when accounting for the larger size of the helical pitch and noting the similarity in the size of the ordered magnetic moment with MnSi. This suggests that the magnetic states in Fe$_{0.7}$Co$_{0.3}$Si, including the helical, conical, and SKX lattice phases, are significantly disordered on a length scale much smaller than $\lambda$. It is surprising, in retrospect, that the helimagnetic and skyrmion lattice ordering survives at all to the extent that the skyrmion lattice state has been observed in both real space images and neutron scattering experiments in this and other $Fe_{1-y}Co_ySi$ compounds with a wide range of $y$. One point should be noted here is that although real space microscopy images show well ordered skyrmion lattice in $Fe_{1-y}Co_ySi$, the neutron scattering experiments on bulk single crystals indicate an orientationally disordered skyrmion lattice. This is primarily due to the reduction in the anisotropic terms that vary as fourth and sixth powers of spin orbit interaction and govern the orientational order of helical and skyrmion lattices [11].

In addition, the line of Brazovskii type first order transition that is clearly visible in $MnSi_{0.962}Al_{0.038}$ and that results from the existence of a fluctuating precursor phase, is not apparent in $Fe_{0.7}Co_{0.3}Si$. Such a difference might result from the smearing of the phase transition due to chemical disorder, the longer length scales in $Fe_{0.7}Co_{0.3}Si$, or a fundamental difference in the paramagnetic-to-helimagnetic transition related to the dynamics of the magnetic moments in two compounds as indicated previously [8,20].

In summary, our data and analysis indicate that the disorder associated with chemical substitution, which is 10 times larger in $Fe_{0.7}Co_{0.3}Si$ than in $MnSi_{0.962}Al_{0.038}$, severely diminishes the entropic changes associated with the entry and exit to the A-phase despite the characteristic length-scales of the skyrmion lattice being orders of magnitude longer than that associated with the disorder. The observation of the A-phase in $Fe_{1-y}Co_ySi$ despite this vast reduction in the entropy change associated with this state, indicates an enormous resiliency of these magnetic structures to substantial disorder.

**Acknowledgments:**

The work at Kennesaw State University was supported by faculty startup grant and the travel support from the department of physics, Kennesaw State University.

This material is based upon work supported by the U.S. Department of Energy under EPSCoR Grant No. DE-SC0012432 with additional support from the Louisiana Board of Regents. A portion of this work was performed at the National High Magnetic Field Laboratory, which is supported by National Science Foundation Cooperative Agreement No. DMR-1644779 and the State of Florida.

# Entropic signatures of the skyrmion lattice phase in MnSi$_{1-x}$Al$_x$ and Fe$_{1-y}$Co$_y$Si.

C. Dhital[1*] and J. F. DiTusa[2]

[1]Department of Physics, Kennesaw State University, Marietta, GA, 30060, USA

[2]Department of Physics and Astronomy, Louisiana State University, Baton Rouge, LA, 70803, USA and The Purdue School of Science, IUPUI, Indianapolis IN, 46202 USA

## 1. Demagnetization correction of magnetic field

The demagnetization correction was performed using the relation $H=H_a-NM$ applicable for a rectangular prism shaped sample[1]. In this relation $H$ is the corrected internal field, $H_a$ is the applied field, $N$ is the demagnetization factor and $M$ is the magnetization.

The MnSi$_{0.962}$Al$_{0.038}$ sample employed for these measurements was nearly rectangular in shape with dimensions of 3 x 2 x 0.9 mm. The magnetic field was applied parallel to the longest side of the sample for all measurements. For this case the demagnetization factor $N$ was calculated to be 0.17 using relation as described in [1]. For Fe$_{0.7}$Co$_{0.3}$Si, the sample dimensions were 1.95 x 1.95 x 1.93 mm. The demagnetization factor $N$ was calculated to be 0.33. For our $M$-$T$ data we report the applied field since the internal field will vary with temperature due to the variation of magnetization with temperature. For all other data reported, we have included estimates of the demagnetization fields using the demagnetization factors stated above.

## 2. Demagnetization correction of ac susceptibility

Both the real and imaginary parts of ac susceptibility were corrected for the effects of demagnetization fields employing the relations [2];

$$\chi' = \frac{\chi'_O - 4\pi N(\chi'^2_O + \chi''^2_O)}{(1-4\pi N\chi'_O)^2 + (4\pi N\chi''_O)^2}, \text{ and } \chi'' = \frac{\chi''_O}{(1-4\pi N\chi'_O)^2 + (4\pi N\chi''_O)^2}.$$

Where $\chi'_O$ and $\chi''_O$ are the observed susceptibilities and $\chi'$ and $\chi''$ are corrected susceptibilities, $N$ is demagnetization factor.

We present the comparison of real part (corrected) of the ac susceptibility, $\chi'$ with the derivative of the entropy with respect to the field, $dS/dH$, for $MnSi_{0.962}Al_{0.038}$ and $Fe_{0.7}Co_{0.3}Si$.

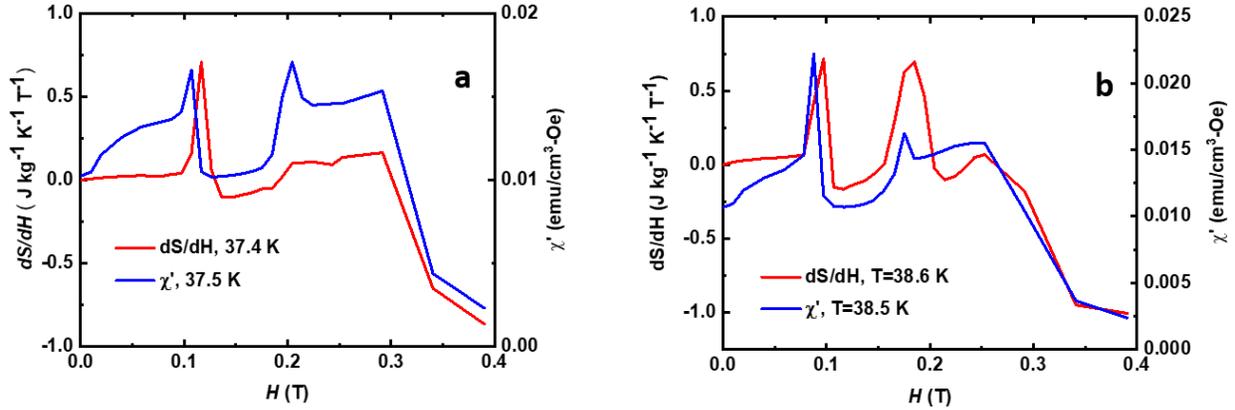

**Fig.** S1. Comparison of real part of ac susceptibility, $\chi'$ and $dS/dH$ for $MnSi_{0.962}Al_{0.038}$ (a) at T~37.5 K (b) at T ~38.5 K.

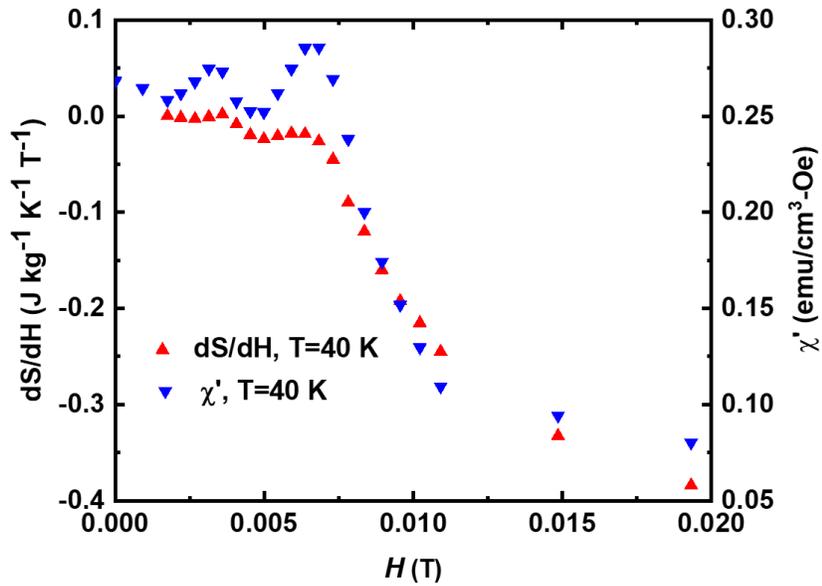

**Fig**. S2. Comparison of real part of ac susceptibility, $\chi'$ and $dS/dH$ for $Fe_{0.7}Co_{0.3}Si$.